\documentclass{article}
\usepackage{spconf,amsmath,graphicx}
\usepackage[table,xcdraw]{xcolor}
\usepackage{booktabs}
\usepackage{cite}
\usepackage{hyperref}
\usepackage{multirow}
\usepackage{diagbox}

\title{Universal Adaptor: Converting Mel-Spectrograms Between Different Configurations for Speech Synthesis}
\name{Fan-Lin Wang$^{1}$, Po-chun Hsu\textsuperscript{*}$^{2}$\thanks{*Equal contribution.}, Da-rong Liu\textsuperscript{*}$^{2}$, Hung-yi Lee$^{1 2}$}
\address{$^1$Department of Electrical Engineering, National Taiwan University, Taiwan\\
$^2$Graduate Institute
of Communication Engineering, National Taiwan University, Taiwan}
\begin{document}
%
\maketitle
\begin{abstract}
Most recent speech synthesis systems are composed of a synthesizer and a vocoder. However, the existing synthesizers and vocoders can only be matched to acoustic features extracted with a specific configuration. Hence, we can't combine arbitrary synthesizers and vocoders together to form a complete system, not to mention apply to a newly developed model. In this paper, we proposed Universal Adaptor, which takes a Mel-spectrogram parametrized by the source configuration and converts it into a Mel-spectrogram parametrized by the target configuration, as long as we feed in the source and the target configurations. Experiments show that the quality of speeches synthesized from our output of Universal Adaptor is comparable to those synthesized from ground truth Mel-spectrogram no matter in single-speaker or multi-speaker scenarios. Moreover, Universal Adaptor can be applied in the recent TTS systems and voice conversion systems without dropping quality.
\end{abstract}
\section{Introduction}
\label{sec:intro}

Neural speech synthesis has achieved remarkable audio qualities recently~\cite{qian2019autovc,lin21b_interspeech,elias2021parallel2, ren2021fastspeech}. 
Most speech synthesis systems comprise two cascaded separated modules: \textbf{synthesizer} and \textbf{vocoder}. 
For instance, in  text-to-speech (TTS), the synthesizer~\cite{shen2018natural, ren2021fastspeech, elias2021parallel, elias2021parallel2} takes text as input and outputs an audio mid-representation.
In voice conversion (VC), the synthesizer~\cite{qian2019autovc,lin21b_interspeech, Chou2019OneshotVC,lin2021fragmentvc} takes a  source speaker's audio as input and outputs a target speaker's audio mid-representation.
Such a representation is typically chosen because it is easier to model than raw audio while preserving enough information to allow faithful inversion back to audio.
In this paper, we follow the most popular works~\cite{wang2017tacotron, shen2018natural, ren2021fastspeech, kalchbrenner2018efficient, Prenger2019WaveglowAF, Kumar2019MelGANGA, Kong2020HiFiGANGA,qian2019autovc,lin21b_interspeech, Chou2019OneshotVC,lin2021fragmentvc} to choose Mel-spectrogram as the mid-representation.
Then in the second module, the vocoder~\cite{oord2016wavenet, kalchbrenner2018efficient, Prenger2019WaveglowAF, yamamoto2020parallel, hsu2020wg, Kumar2019MelGANGA, Kong2020HiFiGANGA, valin2019lpcnet, gao2021extremely} takes the mid-representation as input and outputs the final waveform.
The vocoder is required to be expressive enough to model the raw audio, which has short- and long-term dependencies at different timescales.

Ideally, the development of the synthesizer and the vocoder can be totally disentangled.
For example, if an author proposes a new vocoder, it should be able to combine with each existing synthesizer directly, and most of them can be found with source codes and pretrained models.
However, we find that these pretrained models may be trained on acoustic features extracted with different speech configurations.
For instance, one of the most popular public implementations of Tacotron 2~\footnote{https://github.com/Rayhane-mamah/Tacotron-2} is conditioned on Mel-spectrograms with a hop size of 275 and a frequency range [55, 7600]. In contrast, another implementation~\footnote{https://github.com/NVIDIA/tacotron2} sets the hop size as 256 and the frequency range as [0, 8000].
This forces the author to either train the proposed vocoder according to different corresponding speech configurations or retrain all synthesizers according to the vocoder's configuration.
Both methods take extra time and computing resources, which may be critical for the research groups with limited resources. A similar situation occurred when developing a new synthesizer.

\begin{figure} [t]
  \includegraphics[width=\linewidth]{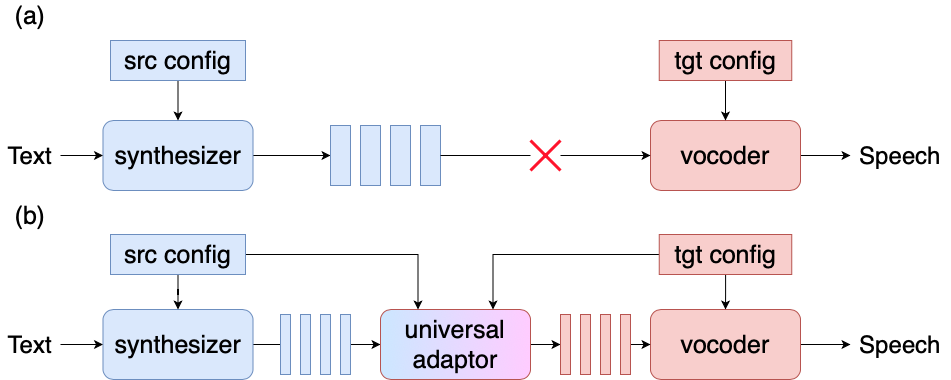}
  \caption{(a) The configurations used for extracting Mel-spectrograms in the synthesizer and the vocoder may be different, limiting the direct cascading of each other.
  (b) The universal vocoder converts Mel-spectrogrms between any two configurations and bridges the gap between the modules.}
  \label{fig:intro}
\end{figure} 

Within the configuration mismatch, some parameters can be fixed by closed-form math conversion, such as elementary arithmetic or log operation. However, other configurations can not be simply conversed. Therefore, a trainable adaptor is required to close the gap. Some studies have similar objectives, such as speech bandwidth expansion (BWE)~\cite{li2018,gupta2019,lin2021,liu2022,andreev2022}. BWE aims to compensate the high-frequency part of a speech signal to increase its resolution. Researchers usually train a deep neural network to perform BWE. Nevertheless, each BWE model can only upsample the signal to its assigned sampling rate, which is not useful when facing arbitrary frequency range, not to mention other time-domain configurations.


To solve the issues, we proposed \textbf{Universal Adaptor}, which can convert a Mel-spectrogram between any two configurations.
With Universal Adaptor, we can cascade any off-the-shelf synthesizer and vocoder even if they are trained with different speech configurations, which is illustrated in figure~\ref{fig:intro}. 
We also demonstrated that Universal Adaptor can be used in any applications involving speech syntheses such as TTS and VC.
Most importantly, there is no performance drop when we cascade models with Universal Adaptor.
Therefore, Universal Adaptor can help determine a vocoder and cascade with any synthesizer for a fair comparison.
In Section \ref{sec:adaptor}, we first described the architecture design of Universal Adaptor.
Then in Section \ref{sec:experiments}, we evaluated the effectiveness of Universal Adaptor on different combinations of synthesizers and vocoders.
Throughout the experiments, all synthesizers and vocoders are collected from pretrained models from open sources.
Except for Universal Adaptor, no additional training is required.
Finally, we concluded our paper in Section \ref{sec:conclusions}. 

\begin{table}[t]
  \caption{All target configurations supported in Universal Adaptor. (a) non-normalizable: parameters without closed-form math conversion; (b) normalizable: parameters with closed-form math conversion.}
  \label{tab:config}
  \centering
  \resizebox{65mm}{!}{\begin{tabular}{lll}
  \toprule\hline
  \bf{Parameter} & \multicolumn{2}{l}{\bf{Value}}                        \\ \hline
  \multicolumn{3}{c}{\textit{(a) non-normalizable}}                      \\ \hline
  wave peak norm & \multicolumn{2}{l}{{[}0.9$\sim$1.0{]}}                \\
  n\_fft         & \multicolumn{2}{l}{{[}1024, 2048{]}}                  \\
  win\_length    & \multicolumn{2}{l}{{[}800, 900, 1024, 1100, 1200{]}}  \\
  hop length     & \multicolumn{2}{l}{{[}window length/4{]}}             \\
  left pad       & \multicolumn{2}{l}{{[}0, (n\_fft-win\_length/4)/2{]}} \\
  right pad      & \multicolumn{2}{l}{{[}0, (n\_fft-win\_length/4)/2{]}} \\
  fmin           & \multicolumn{2}{l}{{[}0, 30, 50, 70, 90{]}}           \\
  fmax           & \multicolumn{2}{l}{{[}7600, 8000, 9500, 11025{]}}     \\ \hline\hline
  \multicolumn{3}{c}{\textit{(b) normalizable}}          \\ \hline
                 &                   & normalizing base  \\
  amp\_to\_db    & {[}True, False{]} & True  \\
  log\_base      & {[}10, 'e'{]}     & 'e'   \\
  log\_factor    & {[}20, 1{]}       & 1     \\
  normalize mel  & {[}True, False{]} & False \\
  ref\_level\_db & {[}0{]}           & 0     \\
  min\_level\_db & {[}-100{]}        & -100  \\ \hline\bottomrule
  \end{tabular}}
\end{table}

\section{Universal Adaptor}
\label{sec:adaptor}
\begin{figure} [t]
  \includegraphics[width=\linewidth]{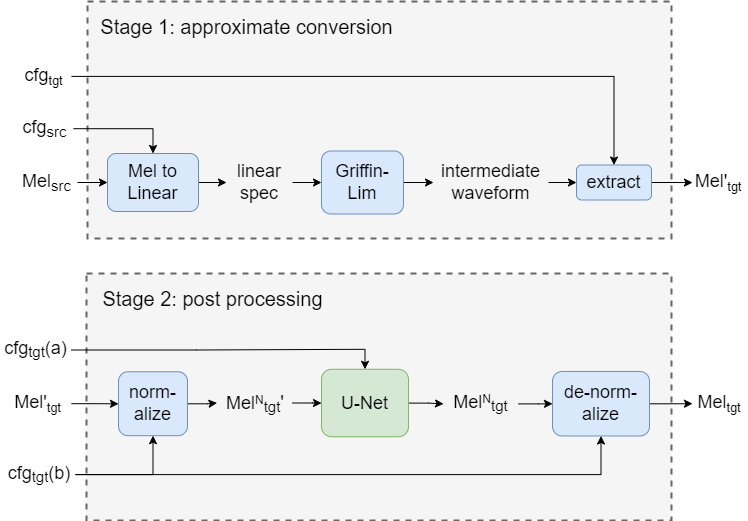}
  \caption{Complete pipeline of Universal Adaptor. $cfg(a)$ is the non-normalizable configuration, and $cfg(b)$ is the normalizable part that has closed-form math conversions.}
  \label{fig:structure}
\end{figure}

Universal Adaptor takes three inputs: the source speech configuration $cfg_{src}$, the target speech configuration $cfg_{tgt}$, and the source Mel-spectrogram $Mel_{src}$ parametrized by $cfg_{src}$. Then, the adaptor generates the target Mel-spectrogram $Mel_{tgt}$ parametrized by $cfg_{tgt}$.
In this paper, we support any arbitrary $cfg_{src}$ and any  $cfg_{tgt}$ listed in Table~\ref{tab:config}, which includes most of the common parameters.
Specifically, all configurations are categorized into two: normalizable and non-normalizable.
The normalizable configurations include those which have simple closed-form math conversion among different choices.
For example, there is a simple closed-form conversion between the Mel-spectrogram that has $log\_base$ 10 and $e$: simply multiplying or dividing by $ln 10$.
On the other hand, the non-normalizable configurations cover the rest of the parameters  that can not simply be conversed.
Universal Adaptor includes two stages as shown in Figure~\ref{fig:structure}, which are described in Section \ref{ssec:arch} and \ref{ssec:unet} respectively. 

\subsection{Stage 1: Approximate conversion}
\label{ssec:arch}
In this stage, we take inputs $cfg_{src}, cfg_{tgt}$ and $Mel_{src}$, and generate the `approximate' target Mel-spectrogram $Mel_{tgt}^{'}$. 
In detail, $Mel_{src}$ is first approximately transformed back to the linear spectrogram, which is done by  multiplication to the pseudo inverse matrix of the Mel-scale filter-bank.
Then we reconstruct the intermediate waveform from the linear spectrogram by  Griffin-Lim algorithm~\cite{griffin1984signal}.
The algorithm is iterated for 32 times.
Afterwards, the waveform is used to generate the $Mel_{tgt}^{'}$ according to $cfg_{tgt}$ with the standard Mel-spectrogram extraction pipeline.
It is worth noting that there are no trainable modules included in this stage.
Because the inversion of the Mel-spectrogram to the linear spectrogram and Griffin-Lim algorithm are only approximations, the reconstructed intermediate waveform and $Mel_{tgt}^{'}$ only have low quality.
Therefore, stage 2 is added to further boost the feature quality.

\begin{figure}
  \centering
  \includegraphics[width=228pt]{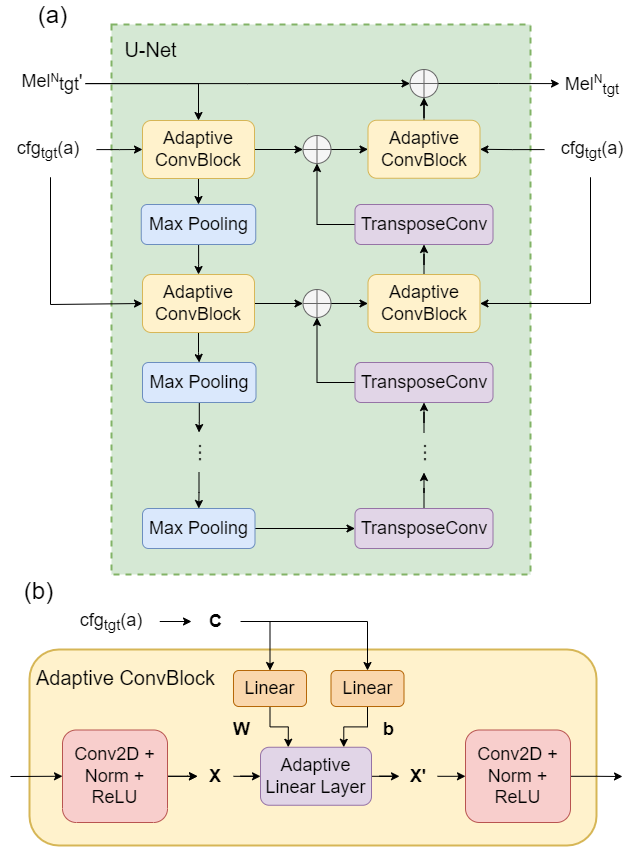}
  \caption{Model architecture of U-Net}
  \label{fig:unet}
\end{figure}

\subsection{Stage 2: Post processing}
\label{ssec:unet}

In this stage, we take $cfg_{tgt}$ and $Mel_{tgt}^{'}$ as input and generate the final target Mel-spectrogram $Mel_{tgt}$. 
The core module of this stage is a U-Net~\cite{ronneberger2015u, wu20p_interspeech, Kameoka2020}. 
Before inputting $Mel_{tgt}^{'}$ into the U-Net, we normalize $Mel_{tgt}^{'}$ according to the normalizable configurations of $cfg_{tgt}$, $cfg_{tgt}(b)$, and denoted the normalized Mel-spectrogram as ${Mel^{N}_{tgt}}'$.
The normalizing base is shown in Table~\ref{tab:config}(b).
Then after obtaining the output of the U-Net, we de-normalize the output, which is $Mel^{N}_{tgt}$, according to $cfg_{tgt}(b)$ and get the final $Mel_{tgt}$.
The reason for normalization is that for each configuration, the Mel-spectrogram is on different scale, and the range of quantity is very large. Therefore, the size of loss will be mostly dependent on the scale of Mel-spectrogram instead of the recovery ability of U-Net. In order to stabilize the training, we only leave U-Net to model the non-normalizable part.

The U-Net module takes ${Mel^{N}_{tgt}}'$ and the non-normalizable configuration of $cfg_{tgt}$, denoted as $cfg_{tgt}(a)$, as input.
The whole architecture is illustrated in Figure~\ref{fig:unet}(a), which consists of an encoder (left side) and a decoder (right side). 
The encoder contains stacks of adaptive ConvBlocks (described in the next paragraph) and Max Pooling layers, which downsample the feature map and increase the number of feature channels; the decoder contains stacks of adaptive ConvBlocks and transposed convolution layers, which upsample the feature map and meanwhile decrease the number of feature channels. 
There are residual connections between the encoder and the decoder at each corresponding block, including the input and the output. 

The adaptive ConvBlock, illustrated in Figure~\ref{fig:unet}(b), is a convolution block that contains an adaptive linear layer~\cite{Perez2018FiLMVR} parameterized by  $cfg_{tgt}(a)$. 
The layer is sandwiched between a typical convolutional neural network structure (2D convolution, batch normalization, and rectified linear unit function).
More specifically for the details, $cfg_{tgt}(a)$ is first encoded as an 8-dimensional vector $\mathbf{C}$. 
In $cfg_{tgt}(a)$, $fmin$ and $fmax$ means the lower bound and upper bound of frequency for Mel basis, and the others are all common arguments in short-time Fourier transform (STFT).
For values with a wider range such as the \emph{hop length} and \emph{win length}, we use the logarithm of the value, while simply use the original value of the others shown in Table~\ref{tab:config}(a).
Then $\mathbf{C}$ determines the weight and the bias of the adaptive linear layer:
\begin{equation} \label{cc}
  \mathbf{W} = Linear(\mathbf{C}), \quad \mathbf{b} = Linear(\mathbf{C}).
\end{equation}
After we obtain the weight and the bias, the vector goes through a nonlinear activation function. Therefore, the complete function of the adaptive linear layer is as follows:
\begin{equation} \label{prelu}
 \mathbf{X'} = PReLU(\mathbf{W} \mathbf{X} + \mathbf{b})
\end{equation}

\begin{table*}[t]
  \caption{Objective evaluation results (MCD, F0-RMSE, V/UV Error) of different models. (inter.: interpolation; Griffin.: Griffin-Lim; adapt.: Universal Adaptor) }
  \label{tab:objective}
  \centering
  \small
  \resizebox{170mm}{!}{\begin{tabular}{lcccccccccccc}
  \toprule\hline
  \multirow{2}{*}{\diagbox{\textbf{Source}}{\textbf{Target}}} & \multicolumn{3}{c}{\textbf{WaveRNN ($cfg 1$)}} & \multicolumn{3}{c}{\textbf{WaveGlow ($cfg 2$)}} & \multicolumn{3}{c}{\textbf{HiFiGAN ($cfg 3$)}} & \multicolumn{3}{c}{\textbf{MelGAN ($cfg 4$)}} \\
        & inter. & Griffin.  & adapt.   & inter. & Griffin.  & adapt.  & inter. & Griffin.  & adapt.   & inter. & Griffin.  & adapt.   \\ \hline
  \multicolumn{13}{c}{\textit{MCD}} \\ \hline
  $cfg 1$ & \cellcolor[rgb]{0.9,0.9,0.9} -  & \cellcolor[rgb]{0.9,0.9,0.9} 9.68 & \cellcolor[rgb]{0.9,0.9,0.9} 9.14 & 48.17 & 17.66 & 15.15 & 49.88 & 10.00 & 7.97 & 23.02 & 13.78 & 10.13 \\
  $cfg 2$ & 49.99 & 32.92 & 12.48 & \cellcolor[rgb]{0.9,0.9,0.9} - & \cellcolor[rgb]{0.9,0.9,0.9} 9.73 & \cellcolor[rgb]{0.9,0.9,0.9}  9.39 & 8.97 & 6.31 & 5.06 & 38.26 & 26.30 & 10.78  \\
  $cfg 3$ & 49.86 & 33.15 & 12.73 & 11.87 & 9.99 & 9.33 & \cellcolor[rgb]{0.9,0.9,0.9} - & \cellcolor[rgb]{0.9,0.9,0.9} 5.95 & \cellcolor[rgb]{0.9,0.9,0.9} 4.78 & 37.46 & 26.24 & 10.56   \\
  $cfg 4$ & 17.55 & 9.96 & 9.04 & 41.21 & 10.97 & 9.73 & 43.94 & 6.33 & 5.08 & \cellcolor[rgb]{0.9,0.9,0.9} - & \cellcolor[rgb]{0.9,0.9,0.9} 8.23 & \cellcolor[rgb]{0.9,0.9,0.9} 6.48 \\ \hline\bottomrule
  \multicolumn{13}{c}{\textit{F0-RMSE}} \\ \hline
  $cfg 1$ & \cellcolor[rgb]{0.9,0.9,0.9} - & \cellcolor[rgb]{0.9,0.9,0.9} 6.36 & \cellcolor[rgb]{0.9,0.9,0.9} 7.45 & 50.18 & 11.63 & 9.28 & 44.63 & 8.94 & 8.55 & 27.98 & 8.15 & 7.20 \\
  $cfg 2$ & 56.32 & 24.77 & 6.19 & \cellcolor[rgb]{0.9,0.9,0.9} - & \cellcolor[rgb]{0.9,0.9,0.9} 7.98 & \cellcolor[rgb]{0.9,0.9,0.9} 6.61 & 9.98 & 6.09 & 4.94 & 26.74 & 7.31 & 5.45  \\
  $cfg 3$ & 58.26 & 19.15 & 7.05 & 7.72 & 8.28 & 7.93 &\cellcolor[rgb]{0.9,0.9,0.9} - & \cellcolor[rgb]{0.9,0.9,0.9} 5.36 & \cellcolor[rgb]{0.9,0.9,0.9} 4.50 & 26.60 & 7.22 & 6.12  \\
  $cfg 4$ & 31.58 & 6.12 & 7.05 & 23.63 & 8.96 & 6.07 & 26.56 & 5.70 & 7.05 & \cellcolor[rgb]{0.9,0.9,0.9} - & \cellcolor[rgb]{0.9,0.9,0.9} 5.11 & \cellcolor[rgb]{0.9,0.9,0.9} 4.83  \\ 
  \hline\bottomrule
  \multicolumn{13}{c}{\textit{V/UV Error}} \\ \hline
  $cfg 1$ & \cellcolor[rgb]{0.9,0.9,0.9} - & \cellcolor[rgb]{0.9,0.9,0.9} 5.93 & \cellcolor[rgb]{0.9,0.9,0.9} 6.14 & 14.37 & 9.31 & 8.63 & 14.52 & 9.70 & 8.85 & 15.62 & 11.37 & 6.17  \\
  $cfg 2$ & 13.38 & 10.34 & 7.67 & \cellcolor[rgb]{0.9,0.9,0.9} - & \cellcolor[rgb]{0.9,0.9,0.9} 6.16 & \cellcolor[rgb]{0.9,0.9,0.9} 5.67 & 7.96 & 7.05 & 5.67 & 10.53 & 7.27 & 5.58  \\
  $cfg 3$ & 12.87 & 11.26 & 7.47 & 8.23 & 5.40 & 6.56 & \cellcolor[rgb]{0.9,0.9,0.9} - & \cellcolor[rgb]{0.9,0.9,0.9} 5.70 & \cellcolor[rgb]{0.9,0.9,0.9} 4.53 & 7.62 & 8.20 & 6.41   \\
  $cfg 4$ & 15.98 & 6.86 & 5.46 & 8.52 & 7.34 & 7.12 & 7.40 & 5.29 & 4.83 & \cellcolor[rgb]{0.9,0.9,0.9} - & \cellcolor[rgb]{0.9,0.9,0.9} 6.19 & \cellcolor[rgb]{0.9,0.9,0.9} 3.90 \\ \hline\bottomrule
  \end{tabular}}
\end{table*}

\begin{table}[t]
  \caption{All the configurations we used in the experiments. Only the important parameters are listed in the table.}
  \label{tab:allconfig}
  \centering
  \resizebox{85mm}{!}{\begin{tabular}{lccccccc}
  \toprule\hline
  \bf{Parameter} & cfg1 & cfg2 & cfg3 &   cfg4 &  cfg5 & cfg6 & cfg7        \\ \hline
  \multicolumn{8}{c}{\textit{(a) non-normalizable}}                      \\ \hline
  wave peak norm & 1.0 & 1.0 & 1.0 & 0.95 &  1.0 & 0.95 & 1.0       \\
  n\_fft         & 2048 & 1024 & 1024 &  1024 & 2048 &  1024 &  465        \\
  win\_length    & 1100 & 1024 & 1024 & 1024 & 1200 & 1024 & 465\\
  hop length     & 275 & 256 & 256 & 256 & 300 & 240 & 160      \\
  left pad       & 0 & 0 & 384 & 384 & 0 & 392 & 0\\
  right pad      & 0 & 0 & 384 & 384 & 0 & 392 & 0\\
  fmin           & 40 & 0 & 0 & 0 & 0 & 0 & 80 \\
  fmax           & 11025 & 8000 & 8000 & 11025 & 12000 & 8000 & 8000\\ \hline\hline
  \multicolumn{8}{c}{\textit{(b) normalizable}}          \\ \hline
  amp\_to\_db    & True & True & True & True & True & True & True\\
  log\_base      & 10  & 'e' & 'e' & 10 & 10 & 'e' & 'e' \\
  log\_factor    & 20  & 1 & 1 & 1 & 20 & 1 & 1  \\
  normalize mel  & True & False & False & False & False & False & False\\
  ref\_level\_db & 0  & - & - & - & - & - & -\\
  min\_level\_db & -100  & - & - & - & - & - & - \\ \hline\bottomrule
  \end{tabular}}
\end{table}



\section{Experiments}
\label{sec:experiments}
\subsection{Datasets}
\label{ssec:datasets}
Our adaptor was trained on LibriTTS~\cite{zen2019libritts}, a multi-speaker English corpus derived from the original materials of LibriSpeech~\cite{panayotov2015librispeech} and often used in speech synthesis tasks~\cite{hsu2019towards, valle2020mellotron, kim2020glow, valle2020flowtron}. We downsampled the utterances to 22kHz for training. As for the test set, we considered a single-speaker and a multi-speaker dataset. The former is the most commonly used single-speaker dataset, LJSpeech~\cite{ljspeech17}, consisting of short audio clips of a single speaker reading non-fiction book passages. The other is the CMU\_ARCTIC databases~\cite{kominek2004arctic}, which were constructed as phonetically balanced, US English speaker databases designed for unit selection speech synthesis research. We chose two male and female speakers for the multi-speaker experiment in Section \ref{ssec:subjective} and all seven speakers for the voice conversion experiment in Section \ref{ssec:application}.

\subsection{Training setup}
\label{ssec:training_setup}
In the training phase, we used the AdamW~\cite{loshchilov2017decoupled} optimizer with default parameters. The learning rate starts from 1e-3 and halves every 50 epochs. U-Net contains 4 layers in the encoder and the decoder respectively, and was trained for 100 epochs with a batch size of 32 on 200-frame long segments. 10\% of the training data were randomly chosen for validation.

Theoretically for training, we should randomly sample a source/target configuration pair for each utterance.
However, Griffin-Lim algorithm produces a computation bottleneck in training.
Therefore, in order to speed up training, we performed Grriffin-lim algorithm beforehand.
Before training, we randomly divided all of the training utterances into 100 subsets and generated a configuration served as $cfg_{src}$ for each subset.
$cfg_{src}$ is fixed throughout the training process.
Each subset is then precomputed into intermediate waveforms, which is illustrated in stage 1 of Figure~\ref{fig:structure}, according to the corresponding $cfg_{src}$. 
During training, we randomly generated 100 configurations in each epoch, and randomly sampled a configuration for each intermediate waveform to serve as $cfg_{tgt}$. 
The ground-truth Mel-spectrogram is computed accordingly from the original waveform.
It is worth mentioning that while producing configurations, including $cfg_{src}$ and $cfg_{tgt}$, we avoided the configurations we used for testing in the following experiments to demonstrate the generalization of the proposed method.

With regards to the loss function, we used L1-loss between $Mel_{tgt}$ illustrated in Figure~\ref{fig:structure} and the ground-truth Mel-spectrogram instead of the commonly used L2-loss.
We found that the error between the $Mel_{tgt}$ and the ground-truth is very small in general. 
If L2-loss is used, the loss will be too small and suffer from gradient vanishing. 
All codes and audio samples will be publicly available online.~\footnote{https://faliwang.github.io/Universal-Adaptor/demo/demo.html}.

\begin{table*}[t]
  \caption{MOS results of different models when using acoustic features from LJSpeech ground truth utterances. The scores are reported with 95\% confidence intervals. In the $\textnormal{Orig.}$ rows, $cfg_{src}$ is same as $cfg_{tgt}$, but the Mel-spectrogram does not go through Universal Adaptor.}
  \label{tab:lj}
  \centering
  \small
  {\begin{tabular}{lccccc}
  \toprule\hline
  \diagbox{\textbf{Source}}{\textbf{Target}} & \textbf{WaveRNN ($cfg 1$)} & \textbf{WaveGlow ($cfg 2$)} & \textbf{HiFiGAN ($cfg 3$)} & \textbf{MelGAN ($cfg 4$)} \\ \hline
  \multicolumn{5}{c}{\textit{LJSpeech (MOS: 4.45$\pm$0.121)}} \\ \hline
  $cfg 1$                & \cellcolor[rgb]{0.9,0.9,0.9}3.89$\pm$0.147 & 2.67$\pm$0.192  & 4.13$\pm$0.143  & 2.69$\pm$0.158  \\
  $cfg 2$                & 3.82$\pm$0.160  & \cellcolor[rgb]{0.9,0.9,0.9}2.86$\pm$0.183 & 4.16$\pm$0.132  & 2.80$\pm$0.163  \\
  $cfg 3$                & 3.77$\pm$0.159  & 2.82$\pm$0.184  & \cellcolor[rgb]{0.9,0.9,0.9}4.22$\pm$0.132 & 2.70$\pm$0.156  \\
  $cfg 4$                & 3.40$\pm$0.152  & 2.78$\pm$0.190  & 4.20$\pm$0.138  & \cellcolor[rgb]{0.9,0.9,0.9}2.83$\pm$0.179 \\ \hline
  Orig.                & \cellcolor[rgb]{0.9,0.9,0.9}3.70$\pm$0.163 & \cellcolor[rgb]{0.9,0.9,0.9}2.97$\pm$0.187 & \cellcolor[rgb]{0.9,0.9,0.9}4.31$\pm$0.136 & \cellcolor[rgb]{0.9,0.9,0.9}2.62$\pm$0.167 \\ 
  \hline\bottomrule
  \end{tabular}}
\end{table*}

\begin{table}
  \caption{MOS results of HiFiGAN trained on VCTK~\cite{yamagishi2019vctk} and tested on acoustic features from CMU\_ARCTIC. The scores are reported with 95\% confidence intervals. In the $\textnormal{Orig.}$ rows, $cfg_{src}$ is same as $cfg_{tgt}$, but the Mel-spectrogram does not go through Universal Adaptor. }
  \label{tab:vctk}
  \centering
  \small
  \begin{tabular}{lc}
  \toprule\hline
  \diagbox{\textbf{Source}}{\textbf{Target}} & \textbf{HiFiGAN ($cfg 3$)} \\ \hline
  $cfg 1$           & 3.55$\pm$0.150  \\
  $cfg 2$           & 3.63$\pm$0.168  \\
  $cfg 3$           & \cellcolor[rgb]{0.9,0.9,0.9}3.34$\pm$0.172 \\
  $cfg 4$           & 3.58$\pm$0.176  \\ \hline
  Orig.           & \cellcolor[rgb]{0.9,0.9,0.9}3.51$\pm$0.149 \\
  Ground Truth    & 3.97$\pm$0.152  \\ 
  \hline\bottomrule
  \end{tabular}
\end{table}

\begin{table*}[t]
  \caption{MOS results of different models when using acoustic features from single-speaker TTS models trained on LJSpeech. The scores are reported with 95\% confidence intervals. In the $\textnormal{Orig.}$ rows, $cfg_{src}$ is same as $cfg_{tgt}$, but the Mel-spectrogram does not go through Universal Adaptor.}
  \label{tab:tts}
  \centering
  \small
  {\begin{tabular}{lccccc}
  \toprule\hline
  \diagbox{\textbf{Source}}{\textbf{Target}} & \textbf{WaveRNN ($cfg 1$)} & \textbf{WaveGlow ($cfg 2$)} & \textbf{HiFiGAN ($cfg 3$)} & \textbf{MelGAN ($cfg 4$)} \\ \hline
  \multicolumn{5}{c}{\textit{Single-speaker TTS}} \\ \hline
  Tacotron ($cfg 1$)     & \cellcolor[rgb]{0.9,0.9,0.9}3.44$\pm$0.172 & 2.97$\pm$0.151  & 3.63$\pm$0.159  & 2.71$\pm$0.161  \\
  Tacotron 2 ($cfg 2$)   & 4.16$\pm$0.127  & \cellcolor[rgb]{0.9,0.9,0.9}3.49$\pm$0.167 & 4.34$\pm$0.111  & 3.30$\pm$0.091  \\
  FastSpeech 2 ($cfg 3$) & 3.47$\pm$0.158  & 3.34$\pm$0.174  & \cellcolor[rgb]{0.9,0.9,0.9}3.68$\pm$0.154 & 2.99$\pm$0.172  \\ \hline
  Orig.                & \cellcolor[rgb]{0.9,0.9,0.9}3.38$\pm$0.144 & \cellcolor[rgb]{0.9,0.9,0.9}3.32$\pm$0.162 & \cellcolor[rgb]{0.9,0.9,0.9}3.35$\pm$0.162 & -               \\ 
  \hline\bottomrule
  \end{tabular}}
\end{table*}

\begin{table*}[t]
  \caption{MOS and similarity results of different models when using acoustic features from voice conversion models trained on VCTK. The scores are reported with 95\% confidence intervals. In the $\textnormal{Orig.}$ rows, $cfg_{src}$ is same as $cfg_{tgt}$, but the Mel-spectrogram does not go through Universal Adaptor.}
  \label{tab:vc}
  \centering
  \small
  {\begin{tabular}{lccc}
  \toprule\hline
  \diagbox{\textbf{Source}}{\textbf{Target}} & \textbf{HiFiGAN ($cfg 3$)} & \textbf{PPG-Voc ($cfg 6$)} & \textbf{S2VC-Voc ($cfg 7$)} \\ \hline
  \multicolumn{4}{c}{\textit{MOS}} \\ \hline
  AdaIN-VC ($cfg 5$) & 3.30$\pm$0.102 & 3.35$\pm$0.099 & 3.26$\pm$0.098 \\
  PPG-VC ($cfg 6$)   & 3.52$\pm$0.096 & \cellcolor[rgb]{0.9,0.9,0.9}3.56$\pm$0.089 & 3.34$\pm$0.094 \\
  S2VC ($cfg 7$)     & 3.51$\pm$0.094 & 3.52$\pm$0.090 & \cellcolor[rgb]{0.9,0.9,0.9}3.32$\pm$0.098 \\ \hline
  Orig.              & -              & \cellcolor[rgb]{0.9,0.9,0.9}3.53$\pm$0.090 & \cellcolor[rgb]{0.9,0.9,0.9}3.45$\pm$0.093 \\ \hline \hline
  \multicolumn{4}{c}{\textit{Similarity}} \\ \hline
  AdaIN-VC ($cfg 5$) & 3.41$\pm$0.126 & 3.34$\pm$0.129 & 3.28$\pm$0.130 \\
  PPG-VC ($cfg 6$)   & 3.42$\pm$0.120 & \cellcolor[rgb]{0.9,0.9,0.9}3.37$\pm$0.125 & 3.40$\pm$0.136 \\
  S2VC ($cfg 7$)     & 3.53$\pm$0.124 & 3.47$\pm$0.128 & \cellcolor[rgb]{0.9,0.9,0.9}3.41$\pm$0.126 \\ \hline
  Orig.              & -              & \cellcolor[rgb]{0.9,0.9,0.9}3.53$\pm$0.124 & \cellcolor[rgb]{0.9,0.9,0.9}3.48$\pm$0.117 \\ 
  \hline\bottomrule
  \end{tabular}}
\end{table*}

\subsection{Configuration pairs}
\label{ssec:configuration}
In our single-speaker experiments, we used four vocoders: WaveRNN \cite{kalchbrenner2018efficient}, WaveGlow \cite{Prenger2019WaveglowAF}, HiFiGAN \cite{Kong2020HiFiGANGA}, and MelGAN \cite{Kumar2019MelGANGA}. These four are matched to each specific configuration. We named their configurations as $cfg 1$, $cfg 2$, $cfg 3$, and $cfg 4$, respectively. Besides, three pretrained synthesizers were adopted in our TTS experiments, and $cfg 1$ is also the configuration for  Tacotron~\cite{wang2017tacotron}; $cfg 2$ is that for Tacotron 2~\cite{shen2018natural}; $cfg 3$ is that for FastSpeech 2~\cite{ren2021fastspeech}.  In the VC experiments, another three pretrained synthesizers were adopted. The configuration of AdaIN-VC~\cite{Chou2019OneshotVC} is denoted by $cfg 5$; that of PPG-VC~\cite{liu2022} is denoted by $cfg 6$; that of S2VC~\cite{lin21b_interspeech} is denoted by $cfg7$. The official repositories provide a vocoder for PPG-VC, PPG-Voc ($cfg6$), and a vocoder for S2VC, S2VC-Voc ($cfg7$). Note that the official AdaIN-VC uses only the Griffin-Lim algorithm to restore the waveform. All of the seven configurations are listed in Table~\ref{tab:allconfig}. In all the tables of experiment results, the row represents $cfg_{src}$ of the input of Universal Adaptor, and the column represents $cfg_{tgt}$ of the output which is also the configuration of the vocoder. 

\subsection{Objective evaluation}
\label{ssec:objective}
For objective evaluation, we picked three aspects to investigate: 1. Mel-ceptral distortion (MCD), which is to measure the difference between two sequences of Mel-cepstra; 2. F0-RMSE, which is the root mean square error of fundamental frequency between two waveforms; 3. V/UV error, which is the error rate of the voiced and the unvoiced flags between the generated and the reference speech. The metrics are calculated in comparison to the reference waveforms synthesized from ground-truth Mel-spectrograms. 

We compared our proposed method (the columns denoted adapt.) to two baselines. The first baseline is closed-form math conversion. We interpolated $Mel_{src}$ to match with the hop length of $cfg_{tgt}$ and rescaled $Mel_{src}$ by the normalizable parameters of $cfg_{tgt}$. Finally, we synthesized the waveforms from them (the columns denoted inter.). The second baseline is Griffin-Lim algorithm. We synthesized waveforms from $Mel_{tgt}'$, which is the  output of the first stage of Universal Adaptor (the columns denoted Griffin.). 

The results are shown in Table~\ref{tab:objective}. Scores with identical $cfg_{src}$ and $cfg_{tgt}$ are shown with gray backgrounds and reported as the top lines for different vocoders. Note that interpolation takes no effect on $Mel_{src}$ when $cfg_{src}$ and $cfg_{tgt}$ are the same. We first see that the scores in $inter.$ are apparently much higher than the other scores, which means pure interpolation is not enough to fix the configuration mismatch. We can hear from the audio samples that the main distortion comes from the frequency bandwidth mismatch. If the band is narrower in $cfg_{src}$ than that in $cfg_{tgt}$, the pitch of the speech will shift higher, and vice versa. It is because the distribution of the frequency band in $Mel_{src}$ is different from that in $Mel_{tgt}$, which can cause a misunderstanding by the vocoder. Moreover, when $cfg_{src}$ and $cfg_{tgt}$ are the same, the errors in $inter.$ should be zero, except for WaveRNN, which clips the Mel-Spectrogram and causes a difference in rescaling, as well as WaveGlow, which inputs random noises. 

On the other hand, the scores in $adapt.$ have an obvious improvement in most of the combinations comparing to $Griffin.$ That is, our second stage can effectively enhance the quality of $Mel_{tgt}'$. Furthermore, for conversions with different \emph{fmax} in source and target configurations, such as $cfg2$ to $cfg1$, the proposed method can restore the high-frequency information of $Mel_{tgt}$ and leads to lower MCD results.

\subsection{Subjective evaluation}
\label{ssec:subjective}
For subjective evaluation, we performed the Mean Opinion Score (MOS) tests.
We randomly chose 15 utterances from the test sets and synthesized the waveforms in all possible configuration combinations. The raters listened to each utterance and rated pleasantness on a five-point scale. Each file was rated by at least 10 different raters. 

In this subsection, there are two parts of experiments, single-speaker experiment on LJSpeech and multi-speaker experiment on CMU\_ARCTIC. In the experiments, the input of Universal Adaptor is the Mel-spectrogram extracted from ground truth waveforms.
It is worth mentioning that in the Orig. rows of Table~\ref{tab:lj} and \ref{tab:vctk}, the Mel-spectrograms do not go through Universal Adaptor. $cfg_{src}$ is matched to $cfg_{tgt}$. The Orig. scores can be the reference of the original performance. In addition, we added the ground truth waveform as a topline. The corresponding scores were written in  $LJSpeech$ row in Table \ref{tab:lj} and in Ground Truth row in Table \ref{tab:vctk}.

The results of the single-speaker experiment are shown in Table~\ref{tab:lj}. Comparing different columns, we can see that HiFiGAN is the best vocoder in these four. Rest of three are WaveRNN, WaveGlow, and MelGAN in order. The results comply with original papers~\cite{Kong2020HiFiGANGA}. Comparing different rows, we can observe that no matter which source configuration we choose, the results are comparable. It proves that Universal Adaptor is effective for configuration conversion without noticeable distortion. Moreover,  observing the grids with gray background leads to that when the input configuration is the same as the output configuration, our adaptor does not corrupt the quality of Mel-spectrograms. Therefore, Universal Adaptor can successfully convert configurations without affecting the vocoder's performance.

On the other hand, the results of multi-speaker experiments are shown in Table \ref{tab:vctk}. The only vocoder we used is HiFiGAN ($cfg3$), which is the only vocoder pretrained on a multi-speaker dataset that is available online among the four vocoders. No matter which source configuration we use, the MOS results are comparable. The results indicate that the proposed Universal Adaptor does not introduce noticeable distortions when converting Mel-spectrograms from multiple speakers.

\subsection{Speech synthesis applications}
\label{ssec:application}
We applied Universal Adaptor to two applications, TTS and VC, to demonstrate the effectiveness in real situations. In the experiments, the inputs of Universal Adaptor are Mel-spectrograms generated by the corresponding synthesizers.

The TTS experiment results are shown in  Table \ref{tab:tts}. By comparing the vocoders in the columns, we obtained the same results as those in the last experiment. When we examine different synthesizers, it is obvious that Tacotron 2 stands out, no matter in which column. Furthermore, the results of the gray grids indicated that Universal Adaptor can even slightly improve the quality of Mel-spectrograms.

The VC experiment results are in Table~\ref{tab:vc}. Besides MOS scores, we  performed a similarity test on a five-point scale. The vocoders we used are the official vocoders along with HiFiGAN pretrained on VCTK. When comparing the MOS scores of vocoders, PPG-Voc performs slightly better than HiFiGAN and much better than S2VC-Voc in each row. It is predictable because PPG-Voc has a similar model architecture to HiFiGAN and S2VC-Voc is similar to WaveRNN. In addition, no matter using which vocoder, PPG-VC can produce more natural speech than S2VC, but S2VC can produce speech more similar to the target speaker than PPG-VC. AdaIN-VC performed the worst in both tests. The results in both experiments indicated that Universal Adaptor can be applied in speech synthesis applications and convert configurations without affecting a model's performance.

\section{Conclusions}
\label{sec:conclusions}
To solve the mismatch of configurations between synthesizers and vocoders, we proposed Universal Adaptor, which includes two stages. The first stage approximately converts the source Mel-spectrogram into the target Mel-spectrogram with poor quality. In the second stage, our module further boosts the quality of the target Mel-spectrogram, which is shown to be effective in objective evaluation. Moreover, the subjective evaluation results revealed that the waveforms synthesized from Universal Adaptor outputs are comparable to those synthesized from ground truth Mel-spectrograms, no matter in single-speaker or multi-speaker scenarios. The result  proves the ability of converting configurations of Universal Adaptor. Universal Adaptor can also be applied in the complete TTS systems and VC systems  without sacrificing performance, verifying the success of Universal Adaptor.





\vfill\pagebreak
\bibliographystyle{IEEEbib}
\bibliography{strings,refs}

\end{document}